\documentclass[fleqn,usenatbib]{mnras}

\DeclareRobustCommand{\VAN}[3]{#2}
\let\VANthebibliography\thebibliography
\def\thebibliography{\DeclareRobustCommand{\VAN}[3]{##3}\VANthebibliography}

\usepackage{graphicx}	
\usepackage{amsmath}	
\usepackage{amssymb}	
\usepackage{footmisc}   

\usepackage{newtxtext,newtxmath}
\usepackage[T1]{fontenc}

\defcitealias{PaperI}{Paper~I}
\DeclareMathOperator{\sech}{sech}
\newcommand{\vlos}{v_\mathrm{los}}
\newcommand{\bfx}{\bmath{x}}
\newcommand{\bfy}{\bmath{y}}
\newcommand{\D}{\mathcal{D}}

\usepackage{xcolor}

\title[The missing radial velocities of \textit{Gaia}]{The missing radial velocities of \textit{Gaia}: a catalogue of Bayesian estimates for DR3}

\author[A. P. Naik et al.]{
Aneesh P. Naik$^{1}$\thanks{E-mail: aneesh.naik@roe.ac.uk}
and Axel Widmark$^{2,3}$
\\
$^{1}$Institute for Astronomy, University of Edinburgh, Royal Observatory, Blackford Hill, Edinburgh EH9 3HJ, UK\\
$^{2}$Stockholm University and The Oskar Klein Centre for Cosmoparticle Physics,  Alba Nova, 10691 Stockholm, Sweden\\
$^{3}$Dark Cosmology Centre, Niels Bohr Institute, University of Copenhagen, Jagtvej 128, 2200 Copenhagen N, Denmark
}

\date{Accepted XXX. Received YYY; in original form ZZZ}

\pubyear{2022}

\begin{document}
\label{firstpage}
\pagerange{\pageref{firstpage}--\pageref{lastpage}}
\maketitle


\begin{abstract}
In an earlier work, we demonstrated the effectiveness of Bayesian neural networks in estimating the missing line-of-sight velocities of \textit{Gaia} stars, and published an accompanying catalogue of blind predictions for the line-of-sight velocities of stars in \textit{Gaia} DR3. These were not merely point predictions, but probability distributions reflecting our state of knowledge about each star. Here, we verify that these predictions were highly accurate: the DR3 measurements were statistically consistent with our prediction distributions, with an approximate error rate of 1.5\%. We use this same technique to produce a publicly available catalogue of predictive probability distributions for the 185 million stars up to a $G$-band magnitude of 17.5 still missing line-of-sight velocities in \emph{Gaia} DR3. Validation tests demonstrate that the predictions are reliable for stars within approximately 7~kpc from the Sun and with distance precisions better than around 20\%. For such stars, the typical prediction uncertainty is 25-30~km/s. We invite the community to use these radial velocities in analyses of stellar kinematics and dynamics, and give an example of such an application.
\end{abstract}


\section{Introduction}

Our understanding of the Milky Way and its stellar dynamics has made great leaps in recent years, and will continue to do so in the near future. Most significantly, the \emph{Gaia} mission \citep{Gaia2016mission} has publicly released astrometric measurements of more than a billion Milky Way stars, yielding unprecedented insights into our Galaxy and its structure, composition and history.

In the \emph{Gaia} catalogue, the vast majority of objects have proper motions but no measurements for the radial (line-of-sight) velocity $\vlos$. The latter measurements are gathered by observing the redshifts of stellar spectra using the Radial Velocity Spectrometer \citep{Cropper2018}, which has a significantly lower limiting magnitude than the astrometer obtaining parallaxes and proper motions. In June 2022, \emph{Gaia} issued its third data release \citep[DR3;][]{Gaia2022DR3}, which increased the number of objects with radial velocities to 33.8 million from the previous 7.2 million in Data Release 2 and Early Data Release 3 (DR2, EDR3; \citealt{Gaia2018DR2, Gaia2021EDR3}; see also \citet{Katz2023} for a description of the DR3 radial velocities). While this is a significant increase, it is still only a small fraction of the 1.47 billion objects with parallax and proper motion measurements. 

This missing velocity component places strong limitations on what can be learned from \emph{Gaia} data, as only five of the six phase space coordinates are available for most stars. In these cases, it is often useful to guess or predict $\vlos$, preferably while also accounting for the uncertainty of such a prediction. This is useful when studying individual objects in the \emph{Gaia} catalogue, as well as for larger scale stellar population analysis. Prediction uncertainties can then be accounted for in statistical inference (e.g., Bayesian marginalisation). For example, \citet{Widmark2022} used predicted radial velocities in this manner when studying the phase-space spiral using the \emph{Gaia} EDR3 proper motion sample.

The six phase space coordinates of \emph{Gaia} stars are correlated with each other via the distribution function (DF) of the local stellar population. For a given star without a radial velocity measurement, a predictive probability distribution is given by conditioning the distribution function on the star's five remaining coordinates. The goal then is to obtain an approximation or representation of this conditional DF, so that one can generate predictions in this manner across all such stars. Supervised deep learning techniques such as neural networks provide one way to achieve this.

The first effort to use neural networks to predict radial velocities from the other five phase-space coordinates was that of \citet{Dropulic2021}, which demonstrated that an artificial neural network can predict $\vlos$ and its associated uncertainty for a population of simulated stars drawn from a mock \textit{Gaia} catalogue \citep[specificially, that of][]{Rybizki2018}. Subsequently, they applied such a network to \textit{Gaia} EDR3 and demonstrated that the technique is capable of identifying substructures in phase space \citep{Dropulic2023}. The deep learning architecture that we employ in this work differs from that of \citet{Dropulic2021, Dropulic2023} in that we use Bayesian neural networks \citep[BNNs;][]{Titterington2004, Goan2020, Jospin2020}. BNNs differ from classical neural networks in that the model parameters (i.e., the network `weights') are not fixed quantities but are instead randomly drawn from probability distributions; the process of training a BNN involves optimising these probability distributions. As such, a BNN represents a Bayesian posterior  probability over the space of network models, rather than just a best-fit point in model space, thus making them less prone to overfitting. The flexibility of BNNs also allow us to produce full posterior probability distributions for each $\vlos$ prediction; these $\vlos$ posterior distributions are flexible in shape, allowing them to be, for example, non-Gaussian, skewed, and even multi-modal.

In \citet[][hereafter Paper I]{PaperI}, we applied BNNs to the \textit{Gaia} EDR3 data set, and produced blind predictions for the \textit{Gaia} DR3 radial velocity data, the latter being released shortly after the publication of our predictions. The first goal of the present article is to provide a comparison of our predictions with the DR3 measurements.

Having thus validated our methodology, we have trained a new BNN model on the DR3 data and used it to generate a novel catalogue of \textit{Gaia} DR3 radial velocity predictions for 185 million stars up to a $G$-band apparent magnitude of $17.5$. The second goal of this article is to present this catalogue and describe its construction and validation. 

Following the publication of this article, our catalogue will be made publicly available via the \emph{Gaia} mirror archive hosted by the Leibniz-Institut f\"ur Astrophysik Potsdam (AIP), where it can be queried alongside the main \emph{Gaia} DR3 catalogue.\footnote{\label{FN:DOI}DOI: \href{https://doi.org/10.17876/gaia/dr.3/110}{\texttt{10.17876/gaia/dr.3/110}}; prior to publication the data will instead be available for bulk download at \url{http://cuillin.roe.ac.uk/~anaik/MissingRVsDR3Catalogue/}} Some guidance for the usage of the catalogue, as well as the trained BNN model, data queries, plotting scripts, and source code are all also made publicly available.\footnote{\url{https://github.com/aneeshnaik/MissingRVsDR3}}

This work is structured as follows. Sec.~\ref{S:BNN} describes our BNN implementation. In Sec.~\ref{S:PreDR3}, we compare the blind predictions of \citetalias{PaperI} with the radial velocity measurements of \emph{Gaia} DR3. In Sec.~\ref{S:DR3Model}, we describe the training of a new BNN model using the DR3 data, focussing particularly on the methodological differences between this work and that of \citetalias{PaperI}. We validate the model using a subset of 6D \emph{Gaia} stars reserved for the purpose in Sec.~\ref{S:6DValidation}. The main result of this work is the catalogue of radial velocity predictions for the 5D stars in \emph{Gaia} DR3, which is presented in Sec.~\ref{S:DR3Catalogue}, before some concluding remarks in Sec.~\ref{S:Conclusions}.

\section{Bayesian Neural Networks}
\label{S:BNN}

The basic objective of our scheme is to derive, for a given \textit{Gaia} star missing a radial velocity measurement, a probability distribution for $\vlos$ given its position $\mathbf{x}$ and its proper motions $\mathbf{\mu}$. Under the assumption that these stars star are drawn independently from the same underlying distribution function as the set of stars $\mathcal{D}$ which do have radial velocity measurements, the latter set can also be used to inform the probability distribution. In other words, the target distribution is the posterior predictive distribution $p(\vlos | \mathbf{x}, \mathbf{\mu}, \mathcal{D})$, an object we describe more quantitatively below. We compute these posterior predictive distributions using BNNs.

This section briefly describes BNNs in general and our implementation in particular. It is intentionally briefer than the corresponding treatment in \citetalias{PaperI} (Sec.~2 in that article), and we refer the reader there or to a review paper such as that of \citet{Jospin2020} for more details. Since the publication of \citetalias{PaperI}, we have made our BNN implementation \citep[wrapped around \textsc{pytorch}][]{Paszke2019} into a publicly available software package: \texttt{banyan}.\footnote{\url{https://github.com/aneeshnaik/banyan}}

The canonical (non-Bayesian) neural network model is a non-linear function which produces outputs $f_\theta(\bfx)$ for inputs $\bfx$. Here, $\theta$ represents the trainable free parameters of the model: in the neural network parlance, the network \emph{weights}. When training the network on some data $\D$ (comprising a series of $\bfx-\bfy$ pairs), the weights $\theta$ are optimised such that the network `learns' the $\bfx-\bfy$ connection. Subsequently, given a new input $\bfx'$, the network output $f_\theta(\bfx')$ serves as a point prediction for the corresponding $\bfy'$.

The issue with the above approach is that it does not provide an obvious way of expressing uncertainty about the predictions. This might be \textit{epistemic} uncertainty stemming from a lack of training data in parts of the data space, or \textit{aleatoric} uncertainty stemming from instrinsic scatter in the $\bfx-\bfy$ connection and/or from measurement errors in the training data. This issue is addressed with BNNs, in which the network weights are promoted from fixed parameters to random variables to be redrawn on each evaluation of the network from probability distributions (assumed 1D Gaussians in our case). The governing parameters $\psi$ -- in our case the means and widths -- of the weight distributions are now the trainable parameters of the network rather than the weights themselves. The aim is then not to find the best-fit weights as in the deterministic approach outlined above, but the optimal parameter set $\hat\psi$ such that the joint distribution of the weights $q_{\psi}(\theta)$ gives an approximation to the posterior distribution: $p(\theta|\D)$. This technique of obtaining a parametric approximation to the posterior is known as \textit{variational inference}, and $q_{\psi}(\theta)$ is referred to as the \textit{variational posterior} \citep{Blei2017}.

Once the network has been thus trained, an output for a given input $\bfx'$ will not be a point prediction for $\bfy'$, but rather a sample from the posterior predictive distribution
\begin{equation}
\label{E:PosteriorPredictive}
    p(\bfy' | \bfx', \D) = \int d\theta p(\bfy' | \bfx', \theta) p(\theta | \D).
\end{equation}
The various sources of uncertainty discussed above are encapsulated in the weight posterior $p(\theta|\D)$: different network weights will govern the network behaviour in different parts of the data space, and the weights pertaining to regions with insufficient data or with large intrinsic scatter will have correspondingly broad posteriors. This uncertainty is then propagated, via the integral above, into the predictions for $\bfy'$.

Our implementation of variational inference uses a novel optimisation procedure. Each optimisation step comprises the following steps:
\begin{enumerate}
    \item $N_s$ random samples are drawn for each network weight from one-dimensional Gaussian distributions parametrized by the current governing parameters $\psi$.
    \item Using the sampled weights $\theta$, $N_s$ predictions $f_{\theta}(\bfx_i)$ are generated by the network for each input data point $\bfx_i$ in the training set
    \item The predictions for each data point are smoothed into a continuous probability distribution $p(\bfy | \bfx_i, \psi)$, using a logistic kernel with adaptive width \citepalias[cf. ][Eq.~3]{PaperI}.
    \item For each data point, the probability distribution computed above is evaluated to give the likelihood of observing the truth $p(\bfy=\bfy_i | \bfx_i, \psi)$.
    \item The total likelihood $p(\D_y | \D_x, \psi)$ is obtained by multiplying $p(\bfy=\bfy_i | \bfx, \psi)$ over all data points.
    \item The \textit{loss} (the optimisation objective to be minimised) is then given by the average negative log likelihood, $-\ln p(\D_y | \D_x, \psi)/N$, where $N$ is the size of the dataset $\D$.
    \item Finally, $\psi$ is updated via a gradient descent step (in the direction of decreasing loss).
\end{enumerate}
In step (i) of this procedure, drawing Gaussian samples for each network weight should be understood as drawing samples from the variational posterior $q_{\psi}(\theta)$: the ensemble of fitted Gaussian distributions is the variational likelihood, and we additionally assume a flat prior on $\theta$. To adopt a different prior $p(\theta)$, one should sample instead from the product of $p(\theta)$ and the Gaussian ensemble.

Post-training, the posterior predictive distributions (Eq.~\ref{E:PosteriorPredictive}) for new data are generated by simply repeating steps (i), (ii), and (iii) of the procedure above, adopting the optimal parameters $\psi=\hat\psi$.

For our present application to the \textit{Gaia} radial velocities, we construct a BNN with five inputs: Galactocentric X/Y/Z; proper motions in right ascension and declination $\mu_\alpha$/$\mu_\delta$, and one output: $\vlos$. Before being fed to the BNN, the data are shifted and rescaled to give means of approximately zero and standard deviations of approximately one along each dimension. Moreover, at the start of each training epoch, the position and velocity of each star is resampled from its error distribution. This procedure is an approximate way to propagate observational uncertainties into the predictions, in addition to the other sources of uncertainty described above.

The precise details of the implementation, e.g., the architecture and training scheme, are discussed in further detail in \citetalias{PaperI}, and any changes made for our present application to DR3 are listed in Sec.~\ref{S:DR3Model}.

\begin{figure*}
    \includegraphics{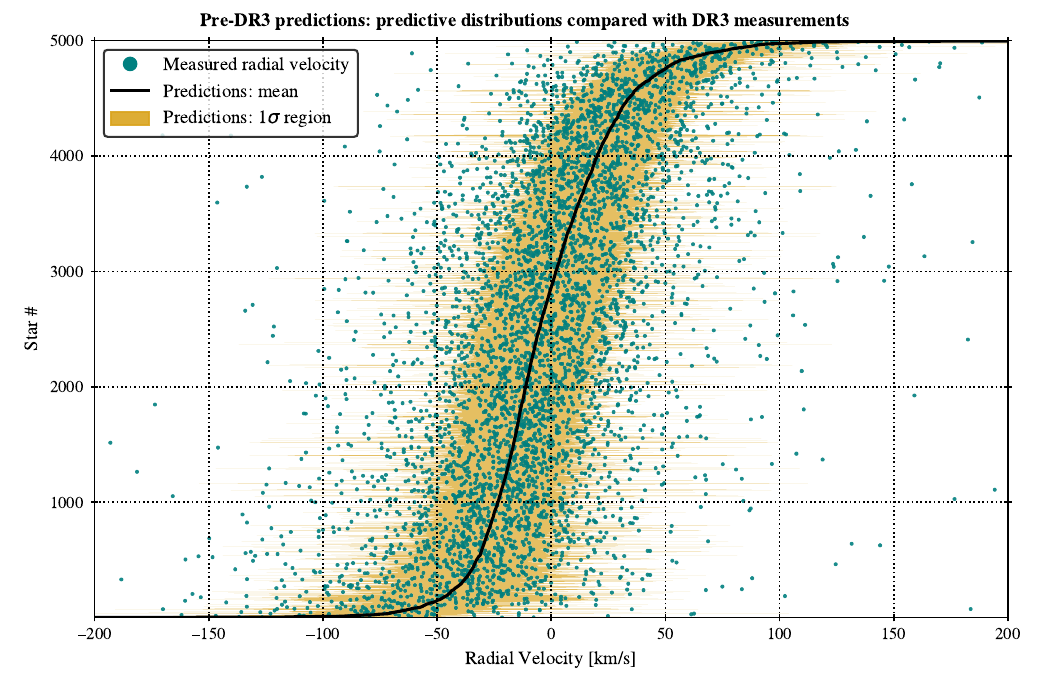}
    \caption{Measured radial velocities and predictive posterior distributions generated by the \emph{Gaia} EDR3-trained BNN for 5000 \emph{Gaia} 6D DR3 stars randomly chosen from the 11 million DR3 stars for which the radial velocity uncertainty is less than 5~km/s and which have entries in the blind prediction catalogue accompanying \citetalias{PaperI}. The solid black line and shaded yellow band indicate the means and standard deviations of the predictive posterior distributions, while the small circular points show the DR3 measurements. The stars are ordered vertically by the means of their prediction distributions. The predictive posteriors are statistically consistent with the true radial velocities, with performance comparable to that achieved in the tests of \citetalias{PaperI}.}
    \label{F:EDR3Posteriors}
\end{figure*}

\begin{figure}
    \includegraphics{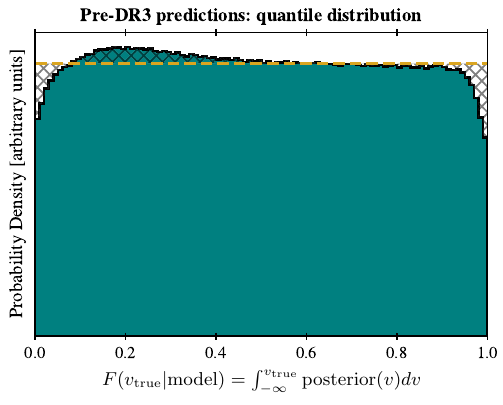}
    \caption{Distribution of `quantiles': positions of the published DR3 radial velocities within the EDR3-derived posterior distributions, for all 11 million DR3 stars for which the radial velocity uncertainty is less than 5~km/s and which have entries in the blind prediction catalogue accompanying \citetalias{PaperI}. The hatched region indicates the area bounded between the quantile distribution and the uniform distribution (the latter indicated with the yellow dashed line), which is used to calculate an approximate error rate (see Eq.~\ref{E:ErrorRate} and surrounding discussion). The distribution is close to uniform indicating good consistency between predictions and measurements (error rate of around 1.5\%), albeit with some evidence of underconfidence.}
    \label{F:EDR3Quantiles}
\end{figure}

\begin{figure}
    \includegraphics{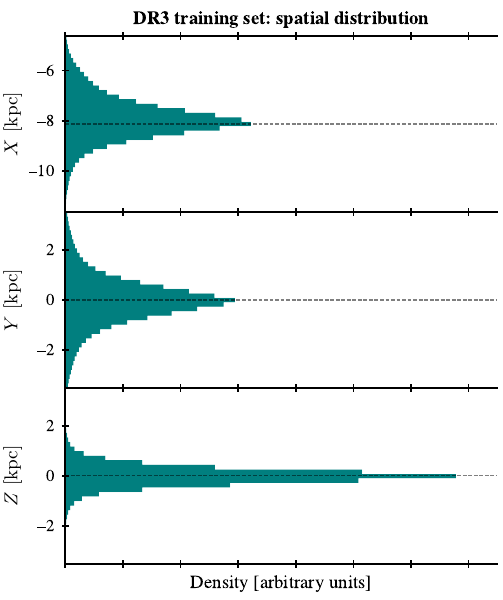}
    \caption{Spatial distribution of the 6D training set. From top to bottom, the three panels respectively show histograms of the stellar density along the Galactocentric $X/Y/Z$ directions. The position of the Sun in these coordinates is assumed to be (8.122, 0, 0.0208) kpc, indicated by the horizontal dashed lines. The characteristic width of the spatial distribution is around 1~kpc in the directions along the disc plane ($X, Y$) and 0.2~kpc in the perpendicular direction ($Z$).}
    \label{F:TrainingSetExtent}
\end{figure}

\section{Pre-DR3 Predictions}
\label{S:PreDR3}

In this section, we compare the blind predictions of \citetalias{PaperI} for the \emph{Gaia} DR3 radial velocities with the published values. Our catalogue provided 250 posterior samples each for 16\,487\,640 stars: all 5D stars in \emph{Gaia} DR2/EDR3 in the magnitude range $6 < G < 14.5$ with accompanying photo-astrometric distance estimates from the \texttt{StarHorse} code \citep{Anders2022}. Of these, we find 14\,581\,884 (88\%) appear in \emph{Gaia} DR3 with measurements for their radial velocities. We further restrict this set to the 10\,953\,995 with $\vlos$ measurement uncertainties less than 5~km/s. We compare our predictions for the radial velocities of these stars with the actual observations.

Figure~\ref{F:EDR3Posteriors} shows, for a randomly chosen subset of 5000 stars, our published prediction distributions and the DR3 measurements. The stars are ordered by the means of the prediction distributions and vertically stacked. The small circular points show the DR3-measured radial velocities, while the solid black line shows the means of the prediction distributions and the yellow shaded region shows one standard deviation above and below the mean (for convenience, we have here neglected that the prediction distributions are in general asymmetric and non-Gaussian; this assumption is dropped in the next figure). This figure gives a first indication that our experiment has been successful: the prediction distributions generally track the measured velocities. Indeed, the appearance of this figure is similar to the analogous figures shown in \citetalias{PaperI} comparing the BNN predictions with mock data and the \emph{Gaia} EDR3 6D test set -- Figs.~2 and 5 respectively in that article\footnote{The analogy between Fig.~\ref{F:EDR3Posteriors} and the figures in \citetalias{PaperI} is not exact because the latter ordered the stars vertically by the \textit{measured} rather than the predicted radial velocity and so gave a misleading impression of discrepancy at the extreme ends of the velocity spectrum, as a natural consequence of `regression to the mean'. This feature vanishes when ordering by mean prediction as in Fig.~\ref{F:EDR3Posteriors}.} -- suggesting that the performance of the BNN procedure on the EDR3 5D sample is comparable to that achieved in tests on the 6D test sample and on mock data.

The success of the predictions is verified more quantitatively in Figure~\ref{F:EDR3Quantiles}, which shows the \textit{quantile distribution} for all 11 million stars (not only a random subset as in the previous figure). This is the distribution of values $F(v_\mathrm{true}|\mathrm{model})$, where for a given star $F$ is the probability -- under the predictive posterior distribution $p(v)$ -- of obtaining a measurement less than the true measurement $v_\mathrm{true}$,
\begin{equation}
    F(v_\mathrm{true}|\mathrm{model}) = \int_{-\infty}^{v_\mathrm{true}}p(v)dv.
\end{equation}
For example, $F=0.5$ indicates that a given radial velocity measurement sits exactly at the median of our prediction distribution, while $F=0$ (1) indicates that the star sits at the extreme low (high) end of the distribution. For a set of prediction distributions that are statistically consistent with the measurement outcomes, one expects a distribution of $F$ values close to a uniform distribution, i.e. $x\%$ of observations are in the bottom $x\%$ of the predictive posteriors, $\forall x$. Indeed, the distribution of $F$ plotted in Fig.~\ref{F:EDR3Quantiles} is very close to the uniform distribution (indicated by the dotted line), indicating that our predictions are generally consistent with the measurements.

There are two small caveats to this success. First, the extremes ($F \approx 0$ and $F \approx 1$) are somewhat underpopulated, suggesting a degree of systematic underconfidence in our predictions: a dearth of data lying at the extremes of probability distributions indicates the distributions are wider than strictly necessary. This underconfidence is a shortcoming, as it means that the precision of our predictions is not optimal. Nonetheless, it is a relatively minor issue as these underpopulated wings are rather small. Moreover, we believe this outcome is better than the opposite case of overconfidence, in which high precisions are achieved at the cost of degraded accuracy.

The second issue apparent in the distribution of $F$ values in Fig.~\ref{F:EDR3Quantiles} is the small bulge at $F \approx 0.2$. This suggests that there is a subset of stars for which the radial velocities are overestimated (typically, these are approaching stars for which the posteriors are incorrectly centred on positive values). We find that this effect does not appear to correlate with any other variables: the bulge in the histogram does not vanish upon restricting the dataset to stars with certain magnitudes, sky positions, proper motions, or distances (or the uncertainties on these quantities). 

Given this quantile distribution, one can estimate an `error rate' as half the total area between the quantile histogram $h$ and the uniform distribution (the hatched region in Fig.~\ref{F:EDR3Quantiles}), 
\begin{equation}
\label{E:ErrorRate}
    E = \frac{1}{2} \int_0^1 | h(x) - 1| dx \approx \frac{1}{2} \sum_i | h_i - 1| \Delta x,
\end{equation}
where the sum is over bins with width $\Delta x$ within the interval [0, 1], and $h_i$ represents the frequency density of stars with $F$ falling in bin $i$. This gives an estimate of the proportion of stars affected by some systematic overestimation or underestimation, with a division by two to account for double-counting. Applying Eq.~\ref{E:ErrorRate} to the quantile histogram displayed in Fig.~\ref{F:EDR3Quantiles}, we find an error rate of around 1.5\%, approximately constant with respect to the number of bins used. Thus, the two systematic effects described above prove to be minor.

We have also repeated this analysis with the pre-DR3 predictions of \citet{Dropulic2023}, taking in particular the 2\,251\,808 stars which appear simultaneously in our catalogue, their catalogue, and the DR3 radial velocity sample (with measurement uncertainty less than 5~km/s). We find that their accuracies are broadly similar to ours, but their prediction uncertainties are generally around 20\% higher. As a result, there is a greater tendency towards underconfidence in their predictions, which show an error rate of 5.5\% versus our 1.7\%.

In summary, our blind predictions for the radial velocities published in \emph{Gaia} DR3 were overall highly successful, with an approximate error rate of 1.5\%. The main purpose of this exercise of generating blind predictions was to validate our technique and advertise it to the community before using it to create a larger catalogue of predictions for radial velocities still missing from DR3. This latter object and its construction is the focus of the remainder of this article.

\section{DR3 Model}
\label{S:DR3Model}

This section describes the construction of a new model trained on the data from \emph{Gaia} DR3. The subsections describe in turn the 6D training data and pre-processing (Sec.~\ref{S:DR3Model:Data}), the BNN implementation (Sec.~\ref{S:DR3Model:BNN}), the training procedure (Sec.~\ref{S:DR3Model:Training}), and the post-training generation of prediction distributions (Sec.~\ref{S:DR3Model:Predictions}). Overall, training the model consumed around 5000 GPU hours.

\subsection{6D Input Data}
\label{S:DR3Model:Data}

There are 33\,653\,049 6D stars in \emph{Gaia} DR3, meaning stars with astrometric measurements as well as radial velocity measurements from the \emph{Gaia} Radial Velocity Spectrometer. We employ these stars to train and test our BNN model.

Whereas in \citetalias{PaperI}, we used stellar distance estimates from the \texttt{StarHorse} code \citep{Anders2022}, here we instead sample distance estimates from the `geometric' posterior described by \citet{BailerJones2021}: a generalised gamma distribution prior with parameters depending on sky location, multiplied by a Gaussian likelihood for the \emph{inverse} distance centred on $\varpi - \varpi_\mathrm{zp}$, where $\varpi$ is the stellar parallax and $\varpi_\mathrm{zp}$ is the parallax zero-point. For the latter quantity, we use the magnitude-, colour- and position-dependent values tabulated by \citet{Lindegren2021} where available, and a uniform value of -0.017~mas otherwise. We sample 10 distance estimates per star, to be used in generating random realisations of the dataset as described in Sec.~\ref{S:DR3Model:Training}. In principle there is a minor inconsistency here: the sky location of a given star influences its estimated distance via both the prior and the likelihood (the latter because of the sky-varying parallax zero-point). As such, to be fully consistent, we should treat sky position and distance as covariant quantities in this random resampling of the dataset. However, because the measurement errors on sky position are very small, it is safe to neglect any such covariance and sample distance and sky position independently.

We perform an 80/20 split on the 6D set to give training and test sets with 26\,922\,439 and 6\,730\,610 stars respectively. We further apply a series of quality cuts to the training set alone, retaining stars for which:
\begin{enumerate}
    \item Radial velocity uncertainty $< 8.5$~km/s.
    \item Uncertainty in both proper motions $< 0.07$~mas/yr.
    \item Fractional uncertainty in distance (estimated as the standard deviation of the 10 distance samples divided by the mean distance sample) $< 5$\%.
    \item $f > 0.5$, where $f$ is the `astrometric fidelity' (an estimate of the quality of a given astrometric solution) as tabulated by \citet{Rybizki2022}.
\end{enumerate} 
Naturally, some of the stars discarded will have reliable measurements which are nevertheless imprecise (e.g., stars which satisfy the last criterion but fail one of the first three). Despite their large uncertainties, such stars would still contain information about the stellar phase space distribution which could further inform our model. Indeed, we shall see later that the cuts restrict the training set to relatively nearby stars, which in turn restricts the range of distances within which our predictions are reliable. However, as discussed in \citetalias{PaperI}, our method for dealing with measurement uncertainties on the training data (resampling on each training epoch) is only robust when these uncertainties are sufficiently small. A more sophisticated treatment of these uncertainties could allow more of the training set to be retained, but we leave the implementation of such a technique for future versions of our model.

Following the cuts, 16\,243\,507 stars remain in the training set. We use these stars to train the model (Sec.~\ref{S:DR3Model:Training}), while the test set remains entirely unseen throughout the training and is subsequently used to validate the trained model (Sec.~\ref{S:6DValidation}).

Figure~\ref{F:TrainingSetExtent} plots the spatial distribution of the training set in three histograms in Galactocentric Cartesian co-ordinates $(X, Y, Z)$, adopting a right-handed system in which the Sun is at $(-8.122, 0, 0.0208)$~kpc \citep{GRAVITY2018, Bennett2019}, as indicated by the vertical lines in the figure. Within the disc-plane, 50/90/99\% of stars in the training set lie within distances $\sqrt{(X - X_\odot)^2 + Y^2}=0.9/2.4/3.9$~kpc, while perpendicular to the disc-plane 50/90/99\% of stars lie within distances $|Z-Z_\odot|=0.2/0.6/1.4$~kpc. These bounding distances would be larger in the absence of the quality cuts made above, which have restricted the training set to the more precise measurements obtained from nearer stars. The spatial limits of our training set are worth bearing in mind when utilising our catalogued predictions: predictions for stars well beyond the range of the training set will necessarily result from a degree of uncertain extrapolation; we explore this further in Sec.~\ref{S:6DValidation}.

\subsection{BNN}
\label{S:DR3Model:BNN}

Here, we merely describe some minor changes in the BNN architecture and computation with respect to that of \citetalias{PaperI}:

\begin{enumerate}
    \item Whereas in \citetalias{PaperI} we trained and generated predictions from a single network, here we train an \emph{ensemble} of 16 networks. Pre-training, each member of the ensemble is assigned a different random initialisation of optimisation parameters (recall that with BNNs, these are not the network weights, but the parameters of the probability distributions from which the weights are drawn). Moreover, during training, each ensemble member is given a different random realisation of the dataset (i.e., different shufflings of the data and different draws from the error distributions, cf. Sec.~\ref{S:DR3Model:Training}). Consequently, each member ends its training at a different local extremum of the optimisation objective. Post-training, we generate a set of $16N$ radial velocity posterior samples for a given star by generating $N$ samples with each ensemble member (Sec.~\ref{S:DR3Model:Predictions}). We find that this ensemble procedure leads to visually smoother, less noisy sample distributions than the single network case, although we note no appreciable change in precision or accuracy.
    \item Because the entire training set can not fit on GPU memory at once, the data are necessarily divided into `batches' which are passed through the network sequentially. On each forward pass through the network, the network weights are randomly sampled from their optimised probability distributions. In the implementation of \citetalias{PaperI}, this would be done once per batch, so all data in a batch would see the same network weights on a given forward pass. However, in some test applications we have found that this can lead to symptoms of underfitting, such as spurious correlations between the output distributions of different inputs. That being said, we have found no evidence of this occurring in our published pre-DR3 predictions (Sec.~\ref{S:PreDR3}). Nonetheless, to avoid this undesirable behaviour in future, we have changed the sampling procedure. Now, the network weights are sampled separately for each data point, so each star in a batch sees different network weights on a given forward pass.
    \item Because the new sampling procedure described above diminishes the danger of underfitting, we have found through extensive testing that we are able to reduce the size of the network architecture without any appreciable decrease in accuracy or precision. We halve the number of hidden layers in the network from 8 to 4, and the number of neurons per layer from 64 to 32. This yields a significant reduction in computational cost, although the total computational cost is still increased with respect to the architecture in \citetalias{PaperI} due to the 16-fold increase in ensemble size.
\end{enumerate}

\subsection{Training}
\label{S:DR3Model:Training}

There are also a few minor changes in the training procedure with respect to that of \citetalias{PaperI}. For clarity, we outline the whole procedure here, highlighting the changes. 

Training a BNN consists of a series of training \textit{epochs} in which the data are fed to the network in a series of batches. At the start of each epoch, the entire dataset is randomly shuffled, and each star has its phase space position randomly drawn from an error distribution: sky positions and proper motions are drawn from a 4D Gaussian centred on the measured values with the published errors and correlations filling the $4\times4$ covariance matrix, radial velocities are drawn from a 1D Gaussian centred on the measured value with width equal to the measurement error, and distances are randomly chosen from the 10 samples from the \citet{BailerJones2021} posterior (cf. Sec.~\ref{S:DR3Model:Data}). This distance sampling differs from that described in \citetalias{PaperI}, in which we sampled distances randomly from a Gaussian centred on the \texttt{StarHorse} value.

From here, training proceeds largely as in \citetalias{PaperI}: the positions are converted to Galactocentric Cartesian coordinates and concatenated with the proper motions to give a length-5 input vector $(X, Y, Z, \mu_\alpha, \mu_\delta)$ alongside a length-1 output vector $(\vlos)$. These inputs / outputs are rescaled by subtracting means of $(-8~\mathrm{kpc}, 0, 0, 0, 0)/(0)$ and converting into units of $(1.5~\mathrm{kpc}, 1.5~\mathrm{kpc}, 0.6~\mathrm{kpc}, 15~\mathrm{mas/yr}, 15~\mathrm{mas/yr})/(40~\mathrm{km/s})$. Note the $Z$ unit has changed from 0.4~kpc in \citetalias{PaperI}, reflecting the larger heights above/below the Galactic disc plane probed by DR3.

The dataset is then split into batches of 400 stars each to be fed to the network sequentially. The batch size has been reduced from 6000 in \citetalias{PaperI} as a result of the increased memory footprint of the altered network parameter sampling procedure described in Sec.~\ref{S:DR3Model:BNN}. At each batch, we perform the optimisation procedure outlined in Sec.~\ref{S:BNN}, generating 256 posterior samples for each star's radial velocity (cf. 250 in \citetalias{PaperI}), scoring them against measurements, and updating the network parameters via a gradient descent step \citep[more specifically, we use the stochastic gradient descent algorithm \textsc{adam};][]{Kingma2015}. We use an initial learning rate of 0.0002. This is significantly reduced from the 0.01 used in \citetalias{PaperI}, compensating for the reduction in batch size (fast learning with small batches can lead to unstable gradient descent). The learning rate is reduced by a factor of 2 once 10 training epochs pass without an appreciable change in the optimisation objective, and the training is ended when the learning rate crosses a threshold value of $5\times 10^{-6}$ (cf. $10^{-5}$ in \citetalias{PaperI}).

The procedure above describes the training of a single network. As noted in Sec.~\ref{S:DR3Model:BNN}, we now train an ensemble of 16 networks rather than a single network.  This procedure is therefore repeated for each member of the ensemble, with different random initialisations of network parameters and different random data shuffling and sampling.

\subsection{Generating Predictions}
\label{S:DR3Model:Predictions}

After undertaking the training described in the previous subsection, the final step is to use the trained model to generate $\vlos$ predictions for the 6D test set (Sec.~\ref{S:6DValidation}) and the 5D set (Sec.~\ref{S:DR3Catalogue}). The procedure in both cases is identical: for each star in the set, 64 posterior samples are generated by each of the 16 trained networks in the ensemble, yielding 1024 posterior samples per star. As in the training stage, each network in the ensemble receives a different error-sampling of the data as an input. These 1024 samples are then converted into a smooth, continuous probability distribution via a logistic ($\sech^2$) smoothing kernel as described in \citetalias{PaperI} (Eq.~2 in that article). For each star, this probability distribution is the posterior predictive distribution for $\vlos$. 

To obtain simpler representations of these probability distributions, we fit them with four-component 1D Gaussian mixtures, so that the posterior for $\vlos$ of a given star is represented as
\begin{equation}
\label{E:GMM}
    p(\vlos) = \sum_{i=1}^4 w_i \frac{\exp{\left(-\displaystyle\frac{(\vlos - \mu_i)^2}{2\sigma_i^2}\right)}}{\sqrt{2\pi\sigma_i^2}},  
\end{equation}
which comprises 12 parameters: the weights $w_{[1,4]}$, the means $\mu_{[1,4]}$, and the variances $\sigma^2_{[1,4]}$ of the Gaussians. Only 11 of these parameters are independent: $\sum_i w_i = 1$ so setting three weights fixes the fourth. Nonetheless, our published catalogue (Sec.~\ref{S:DR3Catalogue}) provides all 12 parameters for simplicity. We find that four Gaussian components are sufficient to give a good fit to the posteriors in all cases, as verified by 1-sample Kolmogorov-Smirnov tests between the BNN posterior samples and the Gaussian mixture CDF. 

As well as facilitating digital storage of the posteriors in a lightweight form, this Gaussian mixture approach is additionally beneficial in that it is subsequently easy to manipulate the distributions, e.g., calculate summary statistics from them, generate further samples from them, and marginalise over them.

\begin{figure*}
    \centering
    \includegraphics{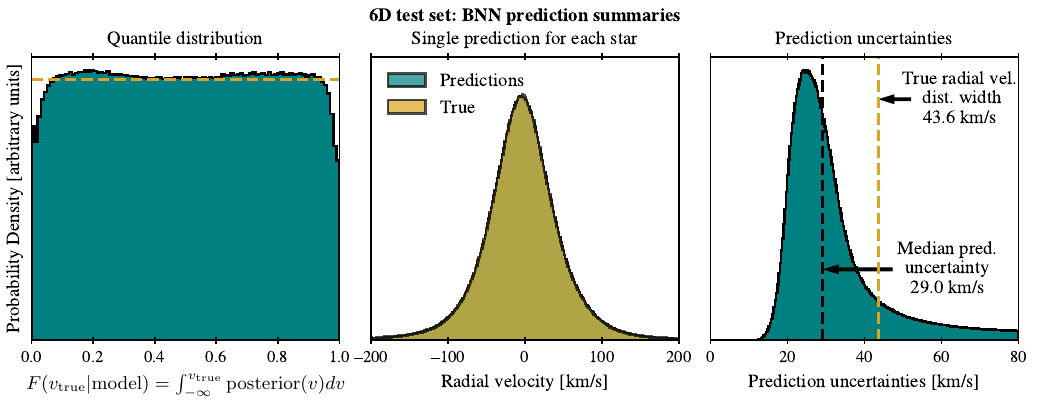}
    \caption{Various summaries of the BNN predictions for the radial velocities of the DR3 6D test set. \emph{Left:} Quantile distribution of measurements within the BNN prediction distributions (analogue of Fig.~\ref{F:EDR3Quantiles}). The yellow dashed line indicates a perfect uniform distribution. \emph{Middle:} histograms of the radial velocities using the measured values (yellow) and a single posterior draw for each star (teal). The two distributions overlap almost exactly. \emph{Right:} Distribution of prediction uncertainties. These are calculated as half the difference between the 84.1\textsuperscript{th} and 15.9\textsuperscript{th} percentile values in the predictive posterior distribution. The vertical dashed line indicates the median uncertainty, and the vertical yellow line indicates the variation in measured $\vlos$ values. The left and middle panels demonstrate the good agreement between the predictions and the measurements, while the right panel demonstrates the predictive power of our technique.}
    \label{F:DR36DTest}
\end{figure*}

\begin{figure}
    \centering
    \includegraphics{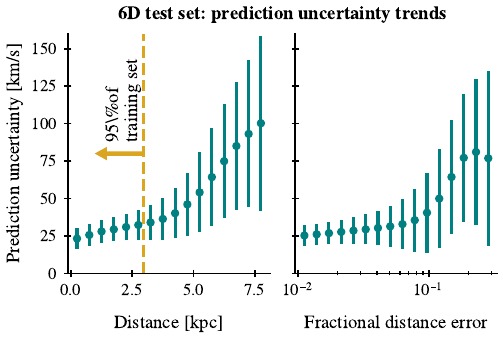}
    \caption{Prediction uncertainties (posterior widths) as a function of distance (\emph{left}) and fractional distance error (\emph{right}). Points and errorbars respectively represent the median and standard deviation of the prediction uncertainties in each bin of the relevant variable. The vertical dashed line in the left-hand panel indicates the heliocentric distance containing 95\% of the training set. This figure demonstrates that the BNN model is able to exercise caution when extrapolating beyond the bounds of its training data, and that it is able to successfully propagate uncertainties on its input data.}
    \label{F:UncertaintyTrends}
\end{figure}

\begin{figure*}
    \includegraphics{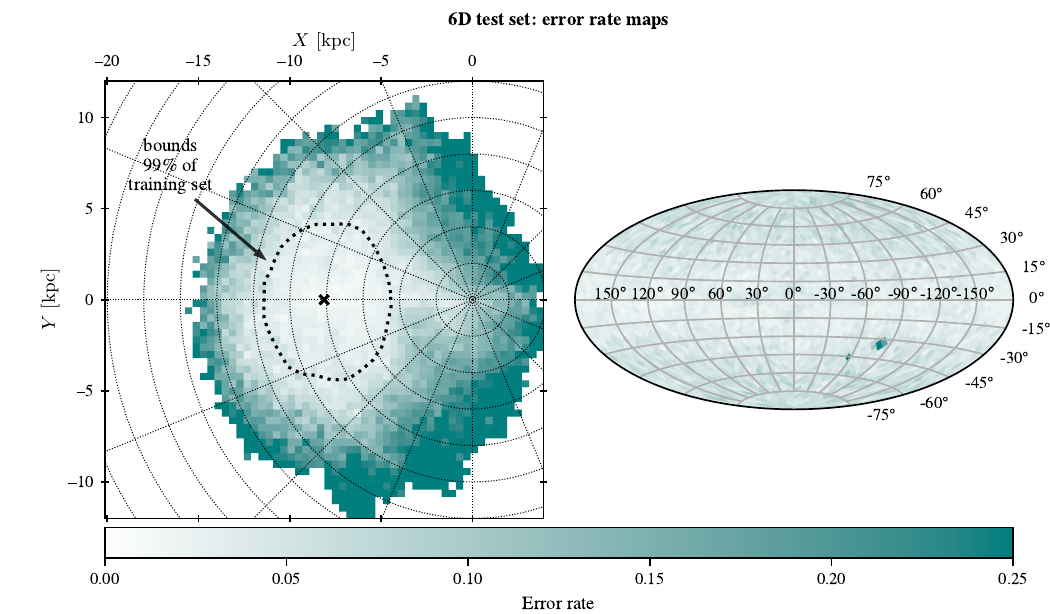}
    \caption{\emph{Left:} Error rates (Eq.~\ref{E:ErrorRate}) calculated for the 6D test set in 2D pixels (width 400~pc) in the Galactic plane. Pixels with fewer than 50 stars are discarded, and the error rate is calculating using 10 quantile bins in each pixel. The cross indicates the position of the Sun, the gridlines give lines of equal Galactocentric radius and azimuth with spacings of 2~kpc and 22.5$^{\circ}$ respectively, and the thick dotted line indicates the (approximately circular) region enclosing 99\% of the training set. \emph{Right:} Error rates calculated in pixels across the sky. The sky is divided into a Healpix grid of order 4, with 3072 sky pixels and a resolution of 3.66$^{\circ}$. The two panels demonstrate that within around 7~kpc and across most of the sky, the error rates are quite low, with problematic regions emerging only at larger distances.}
    \label{F:ErrorMaps}
\end{figure*}

\begin{figure}
    \centering
    \includegraphics{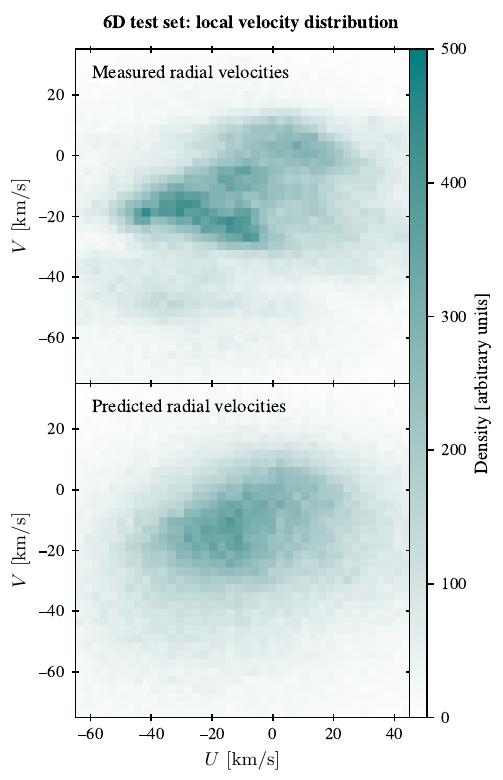}
    \caption{Histogram of velocities parallel to the Galactic plane ($U$ and $V$) for \emph{Gaia} 6D test set stars within a distance of 250~pc. The top panel shows the velocities when using the \emph{Gaia} RVS $\vlos$ measurements, while the bottom panel shows the velocities using predicted $\vlos$ values, drawing a single posterior sample for each star. The complex velocity substructure visible in the top panel is not resolved by our velocity predictions, although the overall shape of the distribution is recovered well.}
    \label{F:LocalVels}
\end{figure}

\section{Model Validation}
\label{S:6DValidation}

In this section, we test and validate the network predictions on the DR3 test set, i.e., the set of 6\,730\,610 6D stars in \emph{Gaia} DR3 not used for network training.

Figure~\ref{F:DR36DTest} shows various summary comparisons between the BNN predictions and the measured radial velocities of the 6D test set. The left-most panel is the analogue of Fig.~\ref{F:EDR3Quantiles}, showing the quantile distribution of the true $\vlos$ measurements within the network's prediction distributions. The distribution is very close to uniform: the error rate calculated with Eq.~\ref{E:ErrorRate} is 1.4\%, similar to the error rate computed in Sec.~\ref{S:PreDR3} for the pre-DR3 predictions. Nonetheless, there is a slight trend visible at the low and high ends, where the histogram value drops to lower values, giving rise to a box shape with rounded corners. This is similar to the feature seen in Fig.~\ref{F:EDR3Quantiles}, and we offer the same interpretation (cf. Sec.~\ref{S:PreDR3}): the far tails of the BNN-generated predictive posterior distributions are, on average, somewhat heavier than necessary; in other words, the network is somewhat overly cautious in terms of the extreme $\vlos$ outliers. While this trend is not ideal, we nonetheless consider it a sign of healthy caution: it is better to produce somewhat under-confident predictions than over-confident ones. The success of the predictions is confirmed by the middle panel, which show histograms of the true $\vlos$ value and randomly drawn prediction posterior realisations; the two distributions overlap almost exactly.

In the rightmost panel of Fig.~\ref{F:DR36DTest}, we plot a histogram of prediction uncertainties (posterior widths). For a given predictive posterior distribution, the uncertainty is defined as half the difference between the 84.1\textsuperscript{th} and 15.9\textsuperscript{th} percentiles. The median prediction uncertainty is 29.0~km/s, although this is a distribution with a long tail to high values (see discussion below) such that the mode is around 24~km/s and there is even a sizable fraction of stars with prediction uncertainties smaller than 20~km/s. 

At first glance, these uncertainties are rather large. Our predictions are of course not a perfect substitute for real measurements (e.g., the median uncertainty in the test set is 3.3~km/s);  the network is learning a representation of the phase space distribution function, which has an \textit{intrinsic} scatter. If a star at a given position and with a given proper motion has a prediction uncertainty of 25~km/s, this is the intrinsic width of the conditional distribution function at that point in position/proper motion space, convolved with any epistemic uncertainty stemming from a lack of training data in the vicinity of that point (this point is discussed further in the Conclusions, Sec.~\ref{S:Conclusions}). Therefore, our prediction uncertainties are not expected to achieve the same precision as the measurements. Instead, we can compare the prediction uncertainties with the uncertainty of a na\"ive guess made in the absence of a model for the distribution function. In the latter scenario, one might simply draw values from a Gaussian distribution with mean zero and width equal to the overall variation in measured $\vlos$. In the test set, the standard deviation of the $\vlos$ measurements is 43.6~km/s. This is significantly larger than the typical uncertainty in our predictions, and this difference quantifies the predictivity afforded by our technique.

It is worth reiterating that the quality cuts used in generating the training set were not applied to the test set (cf. Sec.~\ref{S:DR3Model:Data}). As a consequence, there are stars in the test set at larger distances and with larger measurement uncertainties than those probed by the training set, allowing us to test the extrapolation and error propagation of our technique respectively, by ensuring that the prediction uncertainties for such stars are inflated correspondingly. We begin to see a hint of this behaviour in the histogram of prediction uncertainties in the rightmost panel of Fig.~\ref{F:DR36DTest}, where there are some stars for which the prediction uncertainty is very high; even higher than the variation in measured values (43.6~km/s).

We investigate these effects more directly in Figure~\ref{F:UncertaintyTrends}, which plots the prediction uncertainty against distance and fractional distance error. In each case, the stars of the test set were divided into 16 bins of the relevant variable, and within each bin we calculated the median and the standard deviation of the posterior widths, depicted in the figure as points and error bars respectively. For each star's distance, we used the mean of the 10 distance samples, while the absolute distance error is given by the standard deviation of the 10 samples (so that the fractional distance error plotted is the ratio of the mean to the standard deviation). As a function of distance (left panel), the $\vlos$ uncertainty slowly increases for a few kpc before rising quite quickly beyond roughly 3~kpc to $\sim$100~km/s levels. This change at 3~kpc approximately coincides  with the spatial extent of the training set, and demonstrates that the trained model is capable of detecting input data beyond the range of its training data and exercising commensurate caution. A similar trend is seen in the behaviour of the $\vlos$ uncertainty as a function of fractional distance error (right panel): the posterior widths rise slowly to around 0.1, beyond which they rapidly grow quite large. Uncertainties on the input data are successfully propagated to uncertainties on the predictions. There is a correlation between distance and distance error in the \emph{Gaia} stars, but we have verified that the two trends observed in Fig.~\ref{F:UncertaintyTrends} hold separately: posterior width increases with distance at fixed distance precision and with distance precision at fixed distance, although we have not tested for formal independence.

In Figure~\ref{F:ErrorMaps}, we show how the error rate (Eq.~\ref{E:ErrorRate}) for the 6D test set varies across the Galactic disc and across the sky. In the disc, we split the test set into square bins of width 400~pc in $X$ and $Y$, discarding any bins with fewer than 50 stars. Within each bin, we construct a quantile histogram as in the left panel of Fig.~\ref{F:DR36DTest} and from it calculate an error rate via Eq.~\ref{E:ErrorRate}. Each histogram is calculated in just 10 bins in $F$ to avoid low-number statistics which might artificially inflate the error rate. These error rates across the disc plane are plotted in the left panel of Fig.~\ref{F:ErrorMaps} (more detail is given in Appendix~\ref{A:Quantiles}, which explicitly shows the quantile histograms in different disc locations, rather than just these error rate summaries). There is a circle around the Sun with a radius of approximately 7~kpc within which the error rate is quite low: around a few percent. Recalling that most of the training set stars were situated within 3-4~kpc (a black dotted line enclosing 99\% of the training set is shown in the figure), this again suggests that a degree of extrapolation by a few kpc is `safe'. However, at the edges of this circle and beyond, the error rate starts to grow. This can be ascribed in part to small number statistics at these large distances: small bin counts distort the quantile histograms, leading to inflated error rates. However, this is not the case everywhere. For example, the bin counts are relatively high in the regions around $(X, Y) \approx (0, \pm 4)$ where the error rate is also very high. There is also such a feature at $(X, Y) \approx (-6, -10)$, which corresponds to the Large Magellanic Cloud (LMC). Although the LMC is much farther away, the very large distance errors for its constituent stars in \emph{Gaia} DR3 mean that our distance sampling procedure can `scatter' these stars into regions much closer to the Sun. These regions with large error rates vanish entirely when restricting to only the stars in the test set with small relative distance uncertainties (e.g., less than 10\%), suggesting that imprecise distance estimates are part of the issue in addition to excessive extrapolation.

The right-hand panel of Fig.~\ref{F:ErrorMaps} provides a similar plot, now showing binned error rates as a function of $l$ and $b$, dividing the sky into 3072 equal-area pixels.  For almost the whole sky, the error rate is low (i.e. the quantile histogram is close to uniform), with median and mean values of 4.0\% and 4.3\%. Note that these values are larger than the overall test set error rate of 1.4\%. This discrepancy is primarily due to the low-number noise induced by partitioning the stars into sky pixels, exacerbated by the choice of equal-area bins rather than equal-population bins: the majority of bins lie in sparsely populated regions away from the Galactic disc plane. These issues notwithstanding, the error rate is quite low across the sky, with two key exceptions: the Large and Small Magellanic Clouds. For the reasons discussed in the previous paragraph, the error rates for the stars in these directions is greatly increased. As in the case of the left-hand panel of Fig.~\ref{F:ErrorMaps}, these regions vanish when when restricting to stars with good distance precision.

In Fig.~\ref{F:LocalVels}, we show the $(U,V)$-velocity field of the DR3 6D test set stars in a very local spatial volume (within 250~pc), where $U$ and $V$ are respectively the velocities in the $X$ and $Y$ directions in the solar rest frame. We compare the $(U,V)$-velocity field when using the actual \emph{Gaia} $\vlos$ measurements and randomly drawn samples from the BNN-generated posteriors (taking one sample for each star). This figure is included to illustrate the network's merits and limitations in terms of resolving velocity sub-structures. As shown in the top panel, this very local volume is well resolved in the 6D data and exhibits a complex structure of moving groups, streams, and clusters \citep[cf.][Fig.~22 in that article]{Gaia2018DR2DiscKinematics}. The bottom panel shows the same velocity field, but using a random prediction posterior sample for each star. Comparing the two panels, we see that the small scale structure is washed out, but otherwise the general shape of the 2D velocity distribution is retained. We highlight that this very nearby spatial region constitutes a miniscule part of the training set and the \emph{Gaia} data in general, and that velocity sub-structures are typically not this highly resolved. Hence, resolving such sub-structures is not what the network model is specialised to do. Still, the predicted $\vlos$ values give a robust account of the general structure of the velocity distribution, including higher order statistics like the skew and kurtosis. Moreover, using the \emph{Gaia} 5D sample with our $\vlos$ predictions can still be very useful in terms of discovering and resolving phase-space sub-structures in cases where the proper motions are the more informative quantities. We demonstrate such an application in the following section.

\begin{figure}
    \centering
    \includegraphics{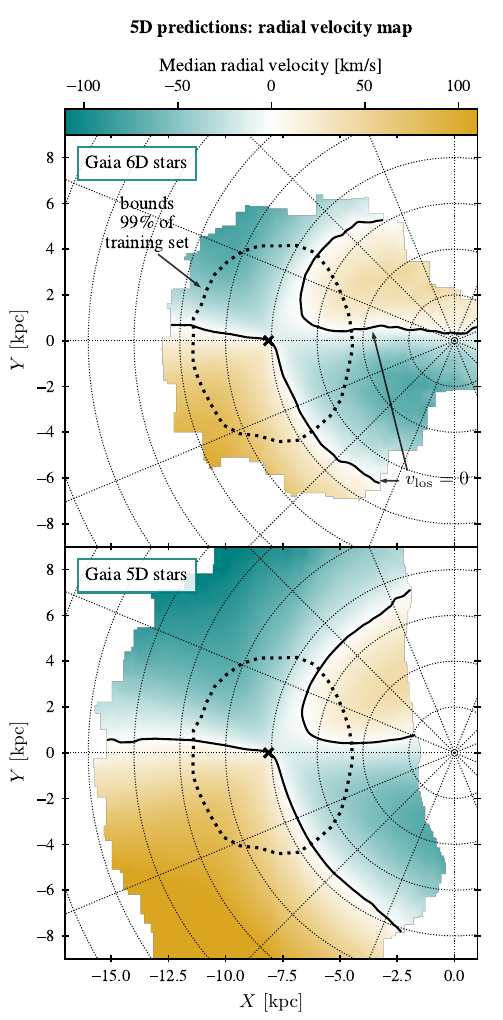}
    \caption{Map of median $\vlos$ across the Galactic disc for stars in the DR3 6D test set (upper panel) and the 5D stars appearing in our BNN-generated catalogue, drawing one posterior sample per star (lower panel). The maps are computed in 50~pc bins in the Galactic disc plane then smoothed with a Gaussian filter of width 150~pc. Some quality cuts are made to the data as described in the main text (Sec.~\ref{S:DR3Catalogue}). In each panel, the cross indicates the position of the Sun, the gridlines give lines of equal Galactocentric radius and azimuth with spacings of 2~kpc and 22.5$^{\circ}$ respectively, and the thick dotted line indicates the (approximately circular) region enclosing 99\% of the training set. Our catalogue enables kinematical analysis with many more stars and to a greater spatial extent than the \emph{Gaia} RVS sample, although excessive extrapolation should be avoided.}
    \label{F:MedianRVMap}
\end{figure}

\begin{figure}
    \includegraphics{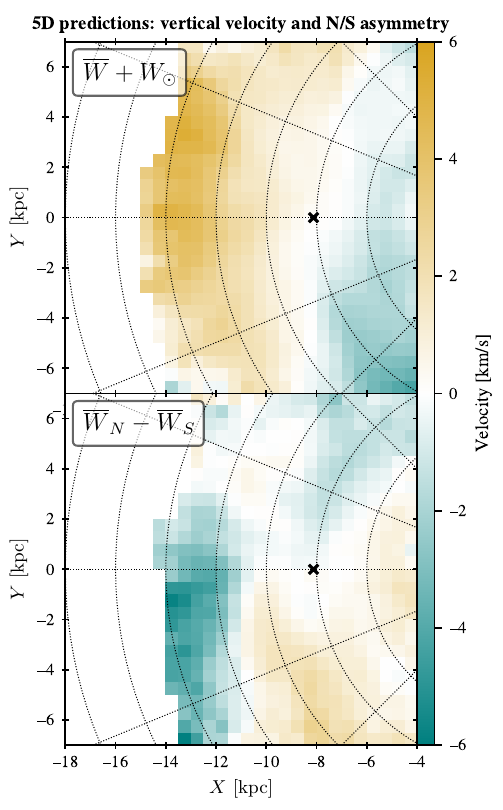}
    \caption{Maps of vertical velocity $W$ across the Galactic disc using 5D stars appearing in our BNN-generated catalogue, drawing one posterior sample per star. In both panels, the sample is limited to $|Z|<600~\mathrm{pc}$ and fractional distance uncertainty better than 20\%. The disc is divided into square pixels with side-length 500~pc and pixels with fewer than 50 stars are masked. For better visibility, each map is then convolved with a 2D Gaussian filter, width 300~pc. The cross indicates the position of the Sun, while the gridlines give lines of equal Galactocentric radius and azimuth, with spacings of 2~kpc and 22.5$^{\circ}$ respectively. \emph{Top panel:} Mean Galactocentric vertical velocity $\overline{W}+W_\odot$ (assuming $W_\odot=7.78$~km/s). \emph{Bottom panel:} Mean vertical velocity difference $W_N - W_S$ between northern and southern Galactic hemispheres. These maps suggest the presence of large-scale bending and breathing disequilibrium modes in the Galactic disc. This serves as an example of a scientific application of our catalogue.}
    \label{F:WMap}
\end{figure}

\section{DR3 Radial Velocity Catalogue}
\label{S:DR3Catalogue}

\begin{table}
	\centering
	\caption{Columns of the BNN-generated $\vlos$ prediction catalogue for \emph{Gaia} DR3. The data in all columns are 32-bit floats except the \texttt{source\_id} column, which comprises 64-bit integers. A more detailed description of the various columns is given in the main text (Sec.~\ref{S:DR3Catalogue}).}
	\label{T:CatalogueColumns}
	\begin{tabular}{ccl} 
		\hline
		Column name & Units &Description\\
		\hline
		\texttt{source\_id}       & -                            & \emph{Gaia} DR3 \texttt{source\_id}\\
		\texttt{w\_\{0,1,2,3\}}   & -                            & Weights of 4 Gaussian components\\
		\texttt{mu\_\{0,1,2,3\}}  & km/s                         & Means of 4 Gaussian components\\
		\texttt{var\_\{0,1,2,3\}} & $(\mathrm{km}/\mathrm{s})^2$ & Variances of 4 Gaussian components\\
		\texttt{q\_050}           & km/s                         & 5\textsuperscript{th} percentile\\
        \texttt{q\_159}           & km/s                         & 15.9\textsuperscript{th} percentile\\
		\texttt{q\_500}           & km/s                         & 50\textsuperscript{th} percentile\\
		\texttt{q\_841}           & km/s                         & 84.1\textsuperscript{th} percentile\\
    	\texttt{q\_950}           & km/s                         & 95\textsuperscript{th} percentile\\	
        \texttt{sample\_mean}     & km/s                         &  Posterior mean\\
        \texttt{sample\_std}      & km/s                         & Posterior standard deviation\\
		\hline
	\end{tabular}
\end{table}

This section describes the main outcome of this work: a BNN-generated catalogue of predictive posterior distributions for the radial velocities of a large number of stars in \emph{Gaia} for which direct measurements are not available. The catalogue is made available via the \emph{Gaia} mirror archive \texttt{gaia.aip.de} hosted by the Leibniz-Institut f\"ur Astrophysik Potsdam (AIP), where it can be queried alongside the main \emph{Gaia} DR3 catalogue (cf. Footnote~\ref{FN:DOI}). The two catalogues can be joined using the \texttt{source\_id} field.

Included in this catalogue are all \emph{Gaia} stars without radial velocities but with 5D astrometry and $G$-band magnitude measurements $G$, up to $G=17.5$. In total, this is 184\,905\,491 stars, 110 million of which have good ($<20\%$) distance precision and are within 7~kpc (96 million within 3~kpc). We imposed this magnitude cut in order to avoid extrapolating too far beyond the training data (the vast majority of the 6D stars in \emph{Gaia} DR3 have $G < 16$) and also to avoid straying beyond the reach of our external validation catalogues (see discussion below). Ultimately there is a degree of arbitrariness in this choice of magnitude cut: the phase space distribution function of the \emph{Gaia} stars has a strong dependence on spatial position and only a weak dependence on absolute magnitude (indeed, we found no improvement in predictivity when including magnitude as a network input), so there will be dim, nearby stars excluded from our catalogue for which the BNN could nonetheless have produced accurate guesses, and bright faraway stars included in the catalogue for which the posteriors are less accurate.

As described in Sec.~\ref{S:DR3Model:Predictions}, we refit the BNN-generated posteriors with four-component Gaussian mixtures to allow a lightweight digital representation and also more easily facilitate `downstream' statistics (e.g., sampling, summary statistics, marginalisation). The final catalogue thus has 20 columns, as enumerated in Table~\ref{T:CatalogueColumns}. The first of these is the \emph{Gaia} DR3 \texttt{source\_id} for each star, an integer uniquely identifying each DR3 star. The next 12 columns are the weights, means and variances of the four Gaussian components (cf. Eq.~\ref{E:GMM}). The remaining columns provide some summary statistics. Columns 14-18 give the 5\textsuperscript{th}, 15.9\textsuperscript{th}, 50\textsuperscript{th} (median), 84.1\textsuperscript{th}, and 95\textsuperscript{th} percentiles $\vlos$ values computed from each posterior, and columns 19-20 are the overall posterior mean and standard deviation.

While the vast majority of stars appearing in this catalogue have never had their radial velocities measured by any instrument, a small subset do have available measurements from other ground-based surveys, allowing us to perform some further validation tests. These are described in Appendix~\ref{A:Validation}.

The remainder of this section describes some preliminary explorations of the catalogue. Figure~\ref{F:MedianRVMap} plots the median $\vlos$, computed in 50~pc bins in the Galactic disc plane then smoothed with a Gaussian filter of width 150~pc, for the DR3 6D test set (upper panel) and for the 5D stars in our catalogue with a single $\vlos$ sampled from the posterior of each star (lower panel). A couple of quality cuts have been made to produce this figure: for both samples, we included only stars with fractional distance uncertainty less than 0.25; for the 5D sample alone, we included only stars with posterior widths less than 80~km/s. This leaves 6\,111\,408 stars in the 6D set and 133\,438\,798 stars in the 5D set. This figure, which is the analogue of Fig.~8 in \citetalias{PaperI}, illustrates that the BNN-predicted velocities reproduce the bulk velocity structure of the Galaxy very well, and our catalogue thus enables kinematical analyses with many more stars and to greater distances than possible with the \emph{Gaia} RVS sample alone. That being said, excessive extrapolation must be avoided. When we relax the quality cuts and include stars with greater distance errors (and by extension, stars at greater distances) we find that at distances $\gtrsim 7$~kpc, the network-predicted Galactic rotation does not quite align with the observed Galactic rotation. This reconfirms our finding with the validation set (Sec.~\ref{S:6DValidation}; in particular Fig.~\ref{F:ErrorMaps}) that extrapolation starts to become somewhat unsafe beyond approximately 7~kpc.

Figure~\ref{F:WMap} provides a second demonstration of the catalogue, mapping the vertical (i.e., perpendicular to the Galactic disc) Galactocentric velocity $W + W_\odot$ in the Galactic plane, including only stars with $|Z|<600~\mathrm{pc}$ and fractional distance uncertainty less than 20\%. The assumed vertical velocity of the Sun is 7.78~km/s \citep{Drimmel2010}. In the most distant regions of this plot, the $\vlos$ predictions are extrapolated beyond the training sample, and so the prediction uncertainties are correspondingly large (cf. Fig.~\ref{F:UncertaintyTrends}). However, the vertical velocity field at such distances is mainly informed by the latitudinal proper motion, and so the contribution of $\vlos$ to the total error budget is subdominant. The upper panel plots the mean $\overline{W} + W_\odot$ in 2D bins in $X$ and $Y$. Here, there is evidence of vertical bulk motions in the disc. The positive slope of $\overline{W} + W_\odot$ with Galactocentric radius suggests a large-scale bending mode and is in agreement with the results of various recent studies mapping the vertical velocity field \citep{WidmarkWidrowNaik2022, Nelson2022, Wang2023}.

The lower panel of Fig.~\ref{F:WMap} plots the difference in mean vertical velocity above and below the disc mid-plane ($\overline{W}_N - \overline{W}_S$). Here, there is evidence of a significant breathing mode. Most notably, in the region at roughly $(X,Y) = (-13,-2)~\mathrm{kpc}$, the disc is undergoing a contraction (i.e. the north and south have a mean velocity towards each other). This structure could be influenced by systematic errors, in either distance or predicted radial velocity. However, it seems improbable that such errors would be able to fully account for the observed bending mode and breathing mode. Furthermore, we begin to see the same structure, albeit with the spatial extent reduced, in both the 6D test set and when making a stronger 10\% distance precision cut.

\section{Conclusions}
\label{S:Conclusions}

In our previous work, \citetalias{PaperI}, we proposed using BNNs to generate estimates for the line-of-sight velocities of \emph{Gaia} stars for which direct measurements are unavailable. The inputs of the model are the five available dimensions (three positions, two proper motions). The outputs are not merely point predictions for the line-of-sight velocities, but full probability distributions which we can understand in a Bayesian sense as predictive posterior distributions. This is the advantage of the BNN approach.

The aims of the present work are two-fold. The first is to follow up on \citetalias{PaperI}, in which we employed a BNN technique on the eve of the third \emph{Gaia} data release (DR3) to generate a set of blind predictions for the radial velocities that would appear in that data release. Here, we compared the predictions with the published measurements, finding very good agreement and an approximate error rate of 1.5\%.

The second purpose of this work is to train a new model with the DR3 measurements, and generate a catalogue of prediction distributions for the radial velocities still missing in DR3. To that end, we used the 6D sample from DR3, comprising around 34 million stars. We split this into a training set and a test set, and applied a series of quality cuts to the training set alone. The training set, finally comprising 16\,243\,507 stars, was then used to optimise the parameters of the model, while the test set was used to validate the model post-training. This validation was successful, yielding an overall error rate of around 1.4\%. There appeared to be a mild degree of underconfidence in the predictions, i.e., the posteriors are wider than necessary, so that a disproportionate number of measurements were sitting near the centres of their prediction distributions. While this behaviour is suboptimal in the sense that we are not achieving the maximum precision attainable, it is still a better outcome than overconfidence. Because of the applied quality cuts, the bulk of the stars of the training set were situated within 3-4~kpc of the Sun. However, we found that the model was capable of a degree of extrapolation, generating accurate posteriors for stars out to around 7~kpc. In general, it achieves this by cautiously inflating the posterior width for stars beyond the range of its training data. However, this behaviour begins to break down beyond this distance, and the predictions grow increasingly less accurate.

Having trained and validated the model, we applied it to all \emph{Gaia} stars without radial velocities but with 5D astrometry and $G$-band magnitude measurements $G$, up to a limit $G=17.5$. This is 184\,905\,491 stars in total, 110 million of which have good ($<20\%$) distance precision and are within 7~kpc (96 million within 3~kpc). For each star, we refitted its BNN-generated posterior distribution with a four-component Gaussian mixture model to give a lightweight representation. The fitted parameters of these Gaussian components, alongside the \emph{Gaia} \texttt{source\_id} of each star and some summary statistics, then form our final catalogue (see Table~\ref{T:CatalogueColumns}), which we have made publicly available.

After constructing this catalogue and validating it against some measurements from ground-based surveys (Appendix~\ref{A:Validation}), we performed some preliminary explorations in order to provide some illustrations of its capabilities. We plotted the median $\vlos$ in the Galactic disc plane, revealing the rotation structure of the disc, out to larger distances than possible with the \emph{Gaia} 6D data alone. We also plotted the mean vertical (perpendicular to the disc) velocity $\overline{W}$ and north-south asymmetry in $\overline{W}$ to large distances in the disc plane, revealing vertical bending and breathing mode disturbances in the disc at large Galactocentric radius (12-14~kpc).

Internally, the network learns a representation of the conditional phase space distribution function $f(\vlos | \mu, \bfx)$ convolved with any epistemic uncertainty arising from the lack of training data. As a result of this, for a given star the prediction distribution is typically quite broad: the typical width under our DR3 model is 25-30~km/s, but this is increased in regions of phase space poorly sampled by the DR3 radial velocity sample.\footnote{As well as epistemic uncertainty, the presence of binary and other multiple-stellar systems (both resolved and unresolved) in the training set will likely have the effect of further smearing the learned distribution function. Such systems might represent as much as half of the stars \citep{Raghavan2010}, but the typical binary orbital speed is only few km/s \citep{ElBadry2021}. Moreover, the orbital speed does not translate directly into a radial velocity measurement bias, but rather sets its upper limit. The velocity measurement is given by an average of several spectrographic observations, and also depends on, for example, the binary orbital plane's orientation. In summary, such biases are a minor effect.}

Conversely, in regions of phase space well sampled by observations, the prediction distribution gives a faithful representation of the conditional phase space distribution function $f(\vlos | \mu, \bfx)$. One could in principle obtain the full distribution function in these regions by multiplying by the 5D density $f(\mu, \bfx)$. This suggests a natural avenue for applications of our catalogue: dynamical mass modelling of the Milky Way disc. For such applications, one must be careful to avoid a double-counting effect: the 6D stars inform the $\vlos$ estimates and are then used again in the mass modelling, and one must ensure that `Bayesian shrinkage' does not occur twice with the same set of observations. A natural approach to using our catalogue which circumvents this issue is \textit{multiple imputation} \citep{Little2014, Gelman2004}, which is formally equivalent to constructing a single joint model for both the missing data and the dynamical parameters of interest $\phi$:
\begin{enumerate}
    \item From our catalogue, draw $N$ random realisations of the predictive posterior for each 5D star, to construct $N$ `complete' datasets.
    \item Conduct the analysis separately on each complete dataset to infer $\phi$.
\end{enumerate}
The variation in the inferred $\phi$ across the $N$ datasets then gives a measure of the `imputation uncertainty': the uncertainty due to the missing data. Multiple imputation is naturally incorporated into Monte Carlo simulation techniques: the $N$ sets of posterior samples for $\phi$ can simply be concatenated to give a final set of posterior samples which marginalises over the missing data.

Beyond dynamics, our catalogue could also be used to better understand the composition and formation history of the Milky Way. As an example of this, \citet{Dropulic2023} used machine-learned $\vlos$ estimates to increase the number of accreted star candidates from the \textit{Gaia}-Sausage/Enceladus merger \citep{Belokurov2018, Helmi2018} by a factor of twenty. Alternatively, one could correlate intrinsic stellar information (e.g., metallicity/age) with the kinematics (e.g., vertical velocity dispersions) of stars in our catalogue to probe the past dynamical heating of the disc.

In all such analyses, valuable information is lost when stars with missing $\vlos$ measurements are discarded. With our catalogue, one can retain these stars in the analysis and marginalise over the missing sixth dimension.

\section*{Acknowledgements}

We are grateful to Bokyoung Kim for assistance with the DESI data, and the \emph{Gaia} mirror archive at the Leibniz-Institut f\"ur Astrophysik Potsdam (AIP) for hosting our catalogue. APN is supported by an Early Career Fellowship from the Leverhulme Trust. AW acknowledges support from the Carlsberg Foundation via a Semper Ardens grant (CF15-0384). This work used the Cirrus UK National Tier-2 HPC Service at EPCC (\url{http://www.cirrus.ac.uk}) funded by the University of Edinburgh and EPSRC (EP/P020267/1). This work has made use of data from the European Space Agency (ESA) mission \textit{Gaia} (\url{https://www.cosmos.esa.int/gaia}), processed by the \textit{Gaia} Data Processing and Analysis Consortium (DPAC, \url{https://www.cosmos.esa.int/web/gaia/dpac/consortium}). Funding for the DPAC has been provided by national institutions, in particular the institutions participating in the \textit{Gaia} Multilateral Agreement. Guoshoujing Telescope (the Large Sky Area Multi-Object Fiber Spectroscopic Telescope LAMOST) is a National Major Scientific Project built by the Chinese Academy of Sciences. Funding for the project has been provided by the National Development and Reform Commission. LAMOST is operated and managed by the National Astronomical Observatories, Chinese Academy of Sciences. For the purpose of open access, the author has applied a Creative Commons Attribution (CC BY) licence to any Author Accepted Manuscript version arising from this submission.

\section*{Data Availability}

All data used in this work are publicly available. The trained BNN model, data queries, plotting scripts, and source code can be accessed at \url{https://github.com/aneeshnaik/MissingRVsDR3}. Following the publication of this article, our catalogue of radial velocity predictions will be made available via the \emph{Gaia} mirror archive hosted by the Leibniz-Institut f\"ur Astrophysik Potsdam (AIP); DOI: \href{https://doi.org/10.17876/gaia/dr.3/110}{\texttt{10.17876/gaia/dr.3/110}}. Prior to publication, the catalogue will instead be available for bulk download at \url{http://cuillin.roe.ac.uk/~anaik/MissingRVsDR3Catalogue/}. The \emph{Gaia} data used to train our model are available at the ESA \emph{Gaia} archive (\url{https://gea.esac.esa.int}) as well as various mirror archives. The LAMOST and DESI data used to validate our catalogue are available for download at \url{http://www.lamost.org/dr8/v2.0/} and \url{https://data.desi.lbl.gov} respectively.

\bibliographystyle{mnras}
\bibliography{library}

\appendix

\section{Test Set Validation: Additional Quantile Distributions}
\label{A:Quantiles}

\begin{figure*}
    \includegraphics{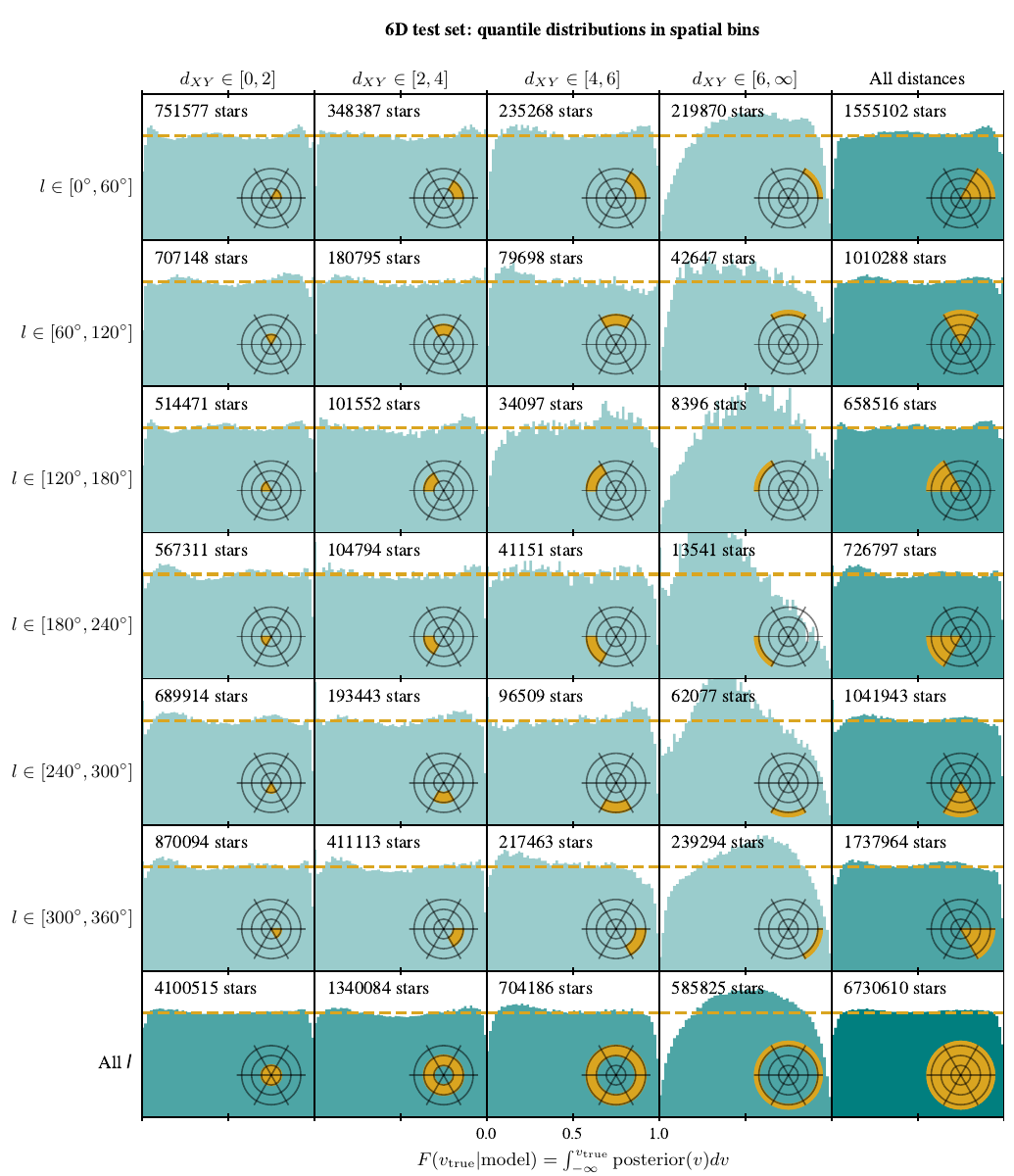}
    \caption{Quantile distributions for the DR3 test set, equivalent to the left panel in Fig.~\ref{F:DR36DTest}, but for different spatial cuts in Galactic longitude (rows) and distance parallel to the Galactic plane (columns), as indicated in the row/column headings and the circular inset diagrams (in which the Galactic centre is to the right and longitude increases anti-clockwise). The rightmost column combines all distances, while the bottom row combines all longitudes, and the bottom right panel combines all longitudes and distances. This figure serves as an elaboration of Figure~\ref{F:ErrorMaps}, plotting the raw quantile distributions across the Galactic disc rather than the integrated error rates.}
    \label{F:6DQuantilesSpatialSplit}
\end{figure*}

In this Appendix, we briefly expand on Fig.~\ref{F:ErrorMaps} in Sec.~\ref{S:6DValidation}, which plotted the variation of the error rate (Eq.~\ref{E:ErrorRate}) across the Galactic disc plane for the 6D test set. While the error rate serves as a useful summary statistic to evaluate the performance of our predictions, some information is necessarily lost when reducing the raw quantile histograms. Thus, it can be instructive to visualise the latter.

In Fig.~\ref{F:6DQuantilesSpatialSplit}, we show quantile histograms for the DR3 6D test set, computed within different spatial bins across the Galactic disc. We divide the $(X,Y)$-plane in terms of Galactic longitude and distance parallel to the Galactic plane (i.e. $d_{XY} \equiv \sqrt{(X-X_\odot)^2+(Y-Y_\odot)^2}$). Overall, the quantile histograms are near box-like, meaning that the network performs well for all these spatial sub-regions in isolation. There are some minor but clear bumps in the quantile histograms, possibly associated with smaller scale phase-space structure that the network has not been capable of learning fully. The most distant (fourth) column, showing stars for which $d_{XY}\geq 6$~kpc, performs the worst: the quantile histograms are more noisy and skewed. In part, this can be ascribed to smaller numbers, but even the more populated angular bins show undesirable features at these distances. For example, there is a visible bump corresponding to the LMC (fourth column, fifth row). These various results mirrors our findings in Sec.~\ref{S:6DValidation}, and highlight the need to use some caution when using predictions that extrapolate beyond the spatial extent of the training set.

\section{External Catalogue Validation}
\label{A:Validation}

\begin{figure}
    \includegraphics{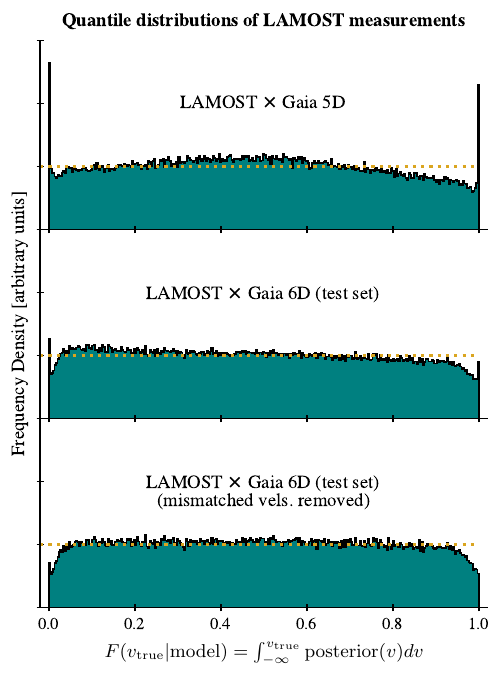}
    \caption{Quantile distribution of LAMOST-measured radial velocities within the BNN-generated predictive posteriors for 191\,598 stars appearing in our 5D catalogue (top panel), 260\,073 stars appearing in our 6D test set (middle panel), and 168\,802 stars appearing in the 6D test set, now excluding stars for which the LAMOST and \emph{Gaia} measurements differ by more than 5~km/s (bottom panel). The top histogram is close to uniform, but with a significant population of outliers. The suggestion of the lower two panels is that these outliers arise due to measurement issues and/or false cross-matches.}
    \label{F:LAMOSTQuantiles}
\end{figure}

\begin{figure}
    \includegraphics{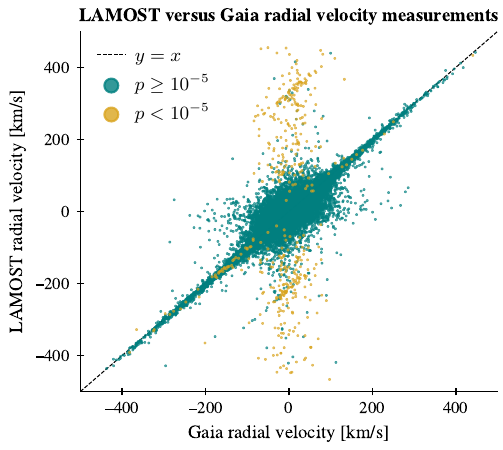}
    \caption{LAMOST $\vlos$ measurements versus \emph{Gaia} $\vlos$ measurements for stars in the 6D test set (circular points). The black dashed line indicates the line $y=x$. Stars with a p-value less than $10^{-5}$ (corresponding to the outliers in Fig~\ref{F:LAMOSTQuantiles}, middle panel) are highlighted in yellow. The LAMOST stars with $\vlos$ falling well outside our prediction distributions appear to have unreliable measurements.}
    \label{F:LAMOSTGaiaComparison}
\end{figure}

\begin{figure}
    \includegraphics{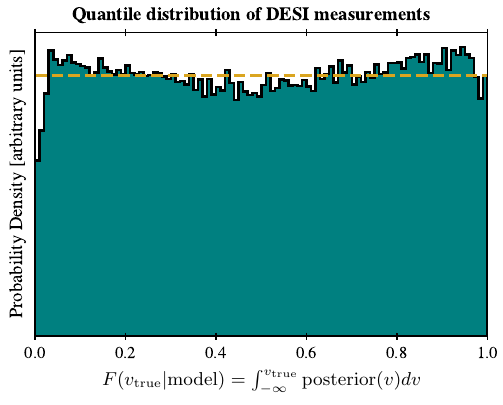}
    \caption{Quantile distribution of DESI-measured radial velocities within the BNN-generated predictive posteriors for 135\,610 DESI stars also appearing in our 5D catalogue. The agreement between the predictions and measurements is very good, albeit with a slight suggestion of overconfidence.}
    \label{F:DESIQuantiles}
\end{figure}

In this section, we describe the results of validating our catalogue of radial velocity estimates against external (i.e., independent of \emph{Gaia}) radial velocity measurements. In particular, we test our catalogue against the radial velocities published in the eighth data release (DR8) of the Large Sky Area Multi-Object Fiber Spectroscopic Telescope \citep[LAMOST;][]{Cui2012, Zhao2012}, and the early data release (EDR) of the Dark Energy Spectroscopic Instrument \citep[DESI;][]{DESI2016, DESI2023}.

\subsection{LAMOST}
\label{A:Validation:LAMOST}

LAMOST \citep{Cui2012, Zhao2012} is a wide-field survey instrument collecting spectroscopic observations of millions of objects in the northern sky: stars within the Milky Way as well as extra-galactic sources.

LAMOST DR8 comprises two components: the `Low-Resolution Spectroscopic Survey' (LRS) and the `Medium-Resolution Spectroscopic Survey' (MRS). With the LRS, we use the sub-catalogue of A, F, G, and K-type stars, while for the MRS we use the main catalogue. In both cases, a large number of observed objects have been matched with \emph{Gaia} DR3 stars. We bulk download both catalogues and apply a series of quality cuts, keeping stars for which:
\begin{itemize}
    \item $\texttt{gaia\_source\_id} \neq -9999$ (there is a match with a \emph{Gaia} DR3 star).
    \item The signal-to-noise ratio > 5. For the LRS, we impose this requirement in in all five bands (\texttt{snru}, \texttt{snrg}, \texttt{snrr}, \texttt{snri}, \texttt{snrz}), while for the MRS we apply it to the combined ratio \texttt{snr}.
    \item There is a reliable radial velocity measurement. For the LRS, we determine this by filtering out stars with radial velocity error set to $\texttt{rv\_err}=-9999$. For the MRS, we check that the flag $\texttt{rv\_br\_flag} = 0$.
\end{itemize}

Within the set of remaining observations, there are a large number of duplicates, i.e., repeat observations of the same star (with the same \emph{Gaia} \texttt{source\_id}). For each distinct \emph{Gaia} \texttt{source\_id}), we combine the various radial velocity measurements into a single variance-weighted average,
\begin{equation}
    \bar{\varv} = \bar{\sigma}^2\sum_i \frac{\varv_i}{\sigma_i^2},
\end{equation}
where the sum is over all observations $i$ of each object, $\varv_i$ and $\sigma_i$ represent individual measurements and errors respectively, and $\bar{\sigma}$ is the combined error on $\bar{\varv}$, given by
\begin{equation}
    \frac{1}{\bar{\sigma}^2} = \sum_i \frac{1}{\sigma_i^2}.
\end{equation}
Note that for the radial velocity measurements $\varv_i$ and errors $\sigma_i$, with the LRS we use the catalogue fields \texttt{rv} and \texttt{rv\_err} respectively, while for the MRS we use the fields \texttt{rv\_br1} and \texttt{rv\_br1\_err}.

With this reduced catalogue, we make one final cut by keeping only stars with precise radial velocity measurements, $\bar{\sigma} < 3.5~\mathrm{km/s}$. Finally, there are 1\,507\,407 stars remaining. Of these, 191\,598 match 5D stars appearing in our catalogue of radial velocity estimates. 

The upper panel of Figure~\ref{F:LAMOSTQuantiles} shows the quantile distribution for the measured velocities of these LAMOST stars within the predictive posteriors generated from our \emph{Gaia} DR3-trained model. Perhaps the most immediately obvious feature is the pair of peaks at the extremes, suggesting a disproportionate number of outliers. We believe that these are not cause for concern, as they coincide with stars for which we have reason to believe that either there is a LAMOST-\emph{Gaia} mismatch or the LAMOST pipeline has estimated $\vlos$ incorrectly, e.g., through the mischaracterisation of spectral lines. To demonstrate that this is plausible, we repeat this exercise of calculating the quantile distribution of the LAMOST stars, but now for the 260\,073 LAMOST sources matching stars in our 6D test set. This is shown in the middle panel of Fig.~\ref{F:LAMOSTQuantiles}. Again, the outlier peaks appear at the extreme ends, albeit with smaller heights than previously. For this set, each star has two $\vlos$ measurements available: the \emph{Gaia} value and the LAMOST value. The bottom panel of Fig.~\ref{F:LAMOSTQuantiles} shows the same quantile distribution replotted after removing the 91\,271 stars for which the two measurements are discrepant by more than 5~km/s (leaving 168\,802 stars). Now, the outlier peaks have vanished. This demonstrates that in the 6D case, the outliers were not a failure of our predictions but an issue with the LAMOST measurements and/or cross-match. 

To reinforce this point, Fig.~\ref{F:LAMOSTGaiaComparison} plots the LAMOST $\vlos$ measurements against the corresponding \emph{Gaia} measurements for the 260\,073 matched stars in the 6D test set. The vast majority of stars lie close to the $y=x$ line, indicating good agreement between the two measurements in general. However, there is a visible minority of stars for which this is not the case. In particular, there is a tall vertical band in the centre of the figure, comprising stars for which \emph{Gaia} measures a relatively small velocity, $|\vlos| \lesssim 40~km/s$, while LAMOST measures much larger velocities of a few 100~km/s. Within this band are two diagonal stripes of stars for which the LAMOST measurements appear to be systematically offset from the Gaia measurements by +325~km/s and -200~km/s in the upper and lower stripe respectively. Highlighted in yellow in the figure are stars occuping the extremities of the quantile plot in the middle panel of Fig.~\ref{F:LAMOSTQuantiles}. In particular, we chose stars with a `p-value' of less than $10^{-5}$ (i.e., $F < 10^{-5}$ for the left tail $F > 1 - 10^{-5}$ for the right). These outlier stars  almost exclusively occupy the vertical band and diagonal stripes in which the LAMOST and \emph{Gaia} measurements are discrepant. We argue that these findings likely hold for our 5D sample (top panel of Fig.~\ref{F:LAMOSTQuantiles}) as well, although this is impossible to truly verify as the \emph{Gaia} measurements are not available by construction. If our assertion is correct, then the outliers can be safely ignored.

Disregarding these peaks at the extremes in the top panel of Fig.~\ref{F:LAMOSTQuantiles}, the predictions for the \emph{Gaia} 5D set are otherwise good: the distribution is close to uniform, albeit with a visible preference for central values, i.e. measurements are disproportionately sitting near the centres of their prediction distributions, suggesting a modicum of under-confidence in the predictions, mirroring our findings when validating the model against the 6D test set (Sec.~\ref{S:6DValidation}). This behaviour seems slightly more pronounced here, reflecting in an increased error rate of 5.1\% (calculated after excluding the first and last quantile bins). This is nonetheless an acceptable degree of deviation, and we can thus conclude that the measured radial velocities of LAMOST are in good agreement with our predictions.

\subsection{DESI}
\label{A:Validation:DESI}

DESI \citep{DESI2016} is a multiobject spectrograph currently undertaking an observational programme with multiple science goals. As well as its main cosmological survey of galaxy redshifts, it will carry out a `Milky Way Survey' \citep{Cooper2023}, ultimately obtaining spectroscopic information (including radial velocities) for several million Milky Way stars. Of these, a few hundred thousand have recently been released as part of the EDR \citep{DESI2023}. In this section, we compare these measurements with our predictions.

From the released data, we bulk download the `Milky Way Survey Catalog of Stellar Parameters' (\texttt{mwsall-pix-fuji.fits}) and apply a series of quality cuts, keeping entries for which:
\begin{itemize}
    \item $\texttt{SOURCE\_ID} \neq 999999$ (there is a match with a \emph{Gaia} DR3 star).
    \item $\texttt{PRIMARY} = \texttt{True}$ (the entry is the primary observation of a given star).
    \item $\texttt{RVS\_WARN} = 0$.
    \item $\texttt{RR\_SPECTYPE} = \texttt{`STAR'}$.
    \item $|\texttt{VRAD}| < 600~\mathrm{km/s}$.
\end{itemize}
The latter three cuts are guided by the recommendations of Koposov et al. (in prep.). Unlike in the LAMOST case (Appendix~\ref{A:Validation:LAMOST}), we make no cut on radial velocity precision, as the vast majority of the remaining stars have sufficiently small uncertainties.

Of the remaining stars, 136\,610 match entries in our catalogue of radial velocity estimates. For these stars, we can compare our predictions with the DESI radial velocity measurements. Fig.~\ref{F:DESIQuantiles} shows the quantile distribution of the DESI-measured radial velocities within the predictive posteriors generated from our \emph{Gaia} DR3-trained model. The distribution shows a slight central dip suggesting a small degree of over-confidence in our predictions, but is otherwise very close to a uniform distribution: the calculated error rate (Eq.~\ref{E:ErrorRate}) is 2.1\%. We therefore conclude that our predictions are in good agreement with the DESI measurements.

\bsp
\label{lastpage}
\end{document}